\documentclass[lettersize,journal]{IEEEtran} 
\usepackage{color,soul}
\usepackage{subcaption}
\usepackage{graphicx}      
\usepackage{comment}
\usepackage{amsmath,amssymb,amsfonts}
\usepackage{tikz}
\usetikzlibrary{arrows.meta, positioning}
\usepackage{tikz-3dplot}
\usepackage[ruled,vlined]{algorithm2e}
\usepackage{amsthm}
\usepackage{cite}

\theoremstyle{plain}

\theoremstyle{remark}

\title{\LARGE \bf
Stealthy Cyber-Attacks on Vehicle Lateral Dynamics:\\A System-Theoretic Analysis
}
\author{Ali Eslami$^{1}$, Jiangbo Yu$^{1}$, and Mohammad Pirani$^{2}$
	\thanks{$^{1}$Ali Eslami ({\tt\small ali.eslami@mcgill.ca}) and Jiangbo Yu ({\tt\small jiangbo.yu@mcgill.ca}) are with the Department of Civil Engineering, Mcgill University, Montreal, Quebec H3A 0G4, Canada.}
		
	\thanks{$^{2}$Mohammad Pirani ({\tt\small mpirani@uottawa.ca}) is with the Department of Mechanical Engineering, University of Ottawa, Ottawa, ON K1N 6N5, Canada. 
		
		}
}
\begin{document}
\maketitle
\thispagestyle{plain}
\pagestyle{plain}
	

\begin{abstract}
	This paper studies the vehicle bicycle model under three classes of stealthy cyber-attacks: replay attacks, zero dynamics attacks, and covert attacks. Using a system-theoretic framework, we analyze the feasibility and impact of these attacks on vehicle lateral dynamics. The investigation considers different measurement configurations, including yaw rate, lateral acceleration, and longitudinal acceleration outputs, to evaluate how sensor selection influences attack detectability and system vulnerability. Each attack class is characterized in terms of required system knowledge, communication access, and impact. The analysis shows that replay attacks remain largely model-agnostic, while zero dynamics attacks are fundamentally constrained by control-oriented design choices, particularly output selection, which can eliminate unstable zero dynamics and limit the attack impact. In contrast, covert attacks, enabled by coordinated actuator and sensor manipulation, allow sustained and stealthy deviation of lateral states when sufficient access and system knowledge are available. The effects of actuator and tire saturation are also examined, revealing attack-dependent impacts on stealthiness and effectiveness. Finally, simulation case studies are conducted by using CarSim–Simulink co-simulation to validate and verify the theoretical results.
\end{abstract}

	

\section{Introduction}
Cyber–Physical Systems (CPS) integrate computation, communication, and control with physical processes and underpin many safety-critical applications, including transportation systems such as road vehicles, aerospace platforms, power grids, and industrial automation \cite{shan2025defense,eslami2023detection,chen2023dynamic}. While this tight integration enables enhanced performance and autonomy, it also exposes CPS to cyber threats that can propagate through networked sensing and actuation and manifest as unsafe physical behavior \cite{huang2025toward}. As a result, ensuring the security and resilience of CPS has become a fundamental research problem. Existing studies broadly address CPS security through attack detection mechanisms \cite{roy2023actuator} and resilient control strategies \cite{li2024event}. 

Among the various attack classes studied in CPS, particular attention has been given to attacks designed to remain stealthy with respect to monitoring schemes. One of the first steps in providing reliable detection and resilient mechanisms is to model the attacks properly. The authors in \cite{teixeira2012attack} investigated several attacks, such as replay attacks, Zero Dynamics Attacks (ZDA), and covert attacks, and provided a mathematical model for them in Networked Control Systems (NCS). In replay attacks \cite{zhao2024replay}, the attacker transmits previously recorded legitimate sensor data. ZDA is another type of attack in which the attacker exploits invariant zeros of the system to drive internal state dynamics while leaving monitored outputs unaffected \cite{eslami2025zero}. Moreover, in covert attacks \cite{eslami2024event}, the attacker coordinates false data injection attacks on both actuator and sensor channels to remove the attack effects from the output of the system and evade detection. Such attacks are particularly concerning in safety-critical CPS, as they can induce substantial performance degradation or instability while remaining undetected. These challenges are further exacerbated in practical CPS deployments, where limited sensing, modeling uncertainties, and physical constraints such as actuator saturation and nonlinearities restrict control authority and complicate both detection and recovery.

Motivated by their safety-critical nature and increasing connectivity, road vehicles have become a prominent focus within the broader CPS security literature \cite{bradley2015optimization}. Prior work on vehicle security has investigated a wide range of attack surfaces, including in-vehicle networks and Electronic Control Units (ECU) \cite{zhang2021cyber}, sensor spoofing and manipulation \cite{dasgupta2022sensor}, and malicious interference with control and estimation modules \cite{liu2025vehicular}. Network-level attacks targeting communication buses such as CAN \cite{fakhfakh2022cybersecurity} have demonstrated the feasibility of injecting or suppressing messages to influence vehicle behavior, while sensor-level attacks on components such as Inertial Measurement Units (IMU), wheel-speed sensors, and positioning systems have shown that corrupted measurements can degrade stability and control performance \cite{liu2019secure}.
However, a system-theoretic analysis of stealthy attacks specifically targeting the lateral control loop remains lacking.

One of the fundamental vehicle dynamics is the lateral dynamics \cite{jiang2018lateral,hu2024vehicles}. Vehicle lateral dynamics govern the motion of a vehicle in the lateral direction and its rotational behavior about the vertical axis, and are therefore fundamental to maintaining stability, maneuverability, and lane-keeping performance \cite{he2024lateral,selman2025lateral}. Key states such as lateral velocity and yaw rate directly influence a vehicle’s ability to follow a desired path and avoid loss of control, particularly during cornering, evasive maneuvers, and low-friction conditions \cite{chen2023dynamic}. As a result, lateral dynamics are central to a wide range of safety-critical functions, including electronic stability control, lane-keeping assistance, and autonomous steering systems. Even small deviations in lateral motion can rapidly escalate into unsafe scenarios, making accurate sensing, estimation, and control of lateral dynamics essential for both human-driven and automated vehicles.

A principled modeling and understanding of attack mechanisms in vehicle lateral dynamics is therefore essential for assessing their impact on safety and control performance. Unlike purely network-level threats, attacks targeting sensing and actuation in lateral control loops interact directly with the physical dynamics that govern stability and path tracking. The effects of such attacks are therefore shaped not only by the attacker’s capabilities, but also by the structure of the vehicle model, the selection of measured outputs, and the limitations imposed by actuators and tire forces. By explicitly modeling the attacks in vehicle's lateral dynamics, both the feasibility and severity of these attacks will become more clear, leading to the design of appropriate detection and mitigation strategies.

In this paper, we investigate stealthy cyber-attacks on vehicle lateral dynamics from a system-theoretic perspective, extending existing CPS security frameworks to this domain. We analyze replay, ZDA, and covert attacks within the lateral control loop, focusing on how output selection, system structure, and physical constraints influence their feasibility and impact. Both linear and nonlinear outputs are considered, along with the effects of actuator and tire saturation.

The main contributions of this paper can therefore be summarized as:
\begin{enumerate}
    \item We systematically model three distinct stealthy attack classes, replay attacks, ZDA, and covert attacks, within the context of vehicle lateral dynamics, characterizing each in terms of required system knowledge, communication channel access, and achievable impact.
    
    \item We demonstrate how control design choices, particularly output selection (yaw rate, lateral acceleration, or longitudinal acceleration), fundamentally alter system vulnerability and the attack design process.
    
    \item  We extend the attack analysis beyond linear output measurements to include nonlinear outputs (specifically, the longitudinal acceleration), deriving the corresponding zero dynamics structure and covert attack coordination requirements for this more complex scenario.
    
    \item We analyze how actuator and tire saturation can influence both the capabilities and stealthiness of each type of attack.
\end{enumerate}

The remainder of this paper is organized as follows. Section II presents the vehicle lateral dynamics model and formulates the problem. Section III provides the theoretical analysis of stealthy attack mechanisms and examines the effects of output selection on the attack feasibility and stealthiness. Section IV investigates the effects of saturation from the perspective of both the attacker and the system. Finally, Section V concludes the paper and discusses directions for future research.

\section{System Model}
\subsection{Vehicle Lateral Dynamics}
The lateral motion of the vehicle can be represented by a two-degree-of-freedom (2-DOF) bicycle model, which captures the lateral velocity \(v_y\) and yaw rate \(r\) of the vehicle. The corresponding state-space model is given by
\begin{equation}
	\dot{x}(t) = A x(t) + B M_z(t) + E \delta(t),
	\label{eq:lateral_dynamics}
\end{equation}
where \(x(t) = [\, v_y \;\; r \,]^{T}\) is the state vector, \(M_z(t)\) is the yaw moment acting on the yaw axis, and \(\delta(t)\) denotes the steering angle input.  
The matrices \(A \in \mathbb{R}^{2\times2}\), \(B \in \mathbb{R}^{2\times1}\), and \(E \in \mathbb{R}^{2\times1}\) are defined as
\begin{equation}
	A = 
	\begin{bmatrix}
		a_{11} & a_{12} \\
		a_{21} & a_{22}
	\end{bmatrix}, \qquad
	B = 
	\begin{bmatrix}
		0 \\ b_2
	\end{bmatrix}, \qquad
	E =
	\begin{bmatrix}
		e_1 \\ e_2
	\end{bmatrix},
	\label{eq:ABC}
\end{equation}
with
	$a_{11} = -\frac{2(C_f + C_r)}{v_x m}, 
	a_{12} = \frac{2(b C_r - a C_f)}{v_x m} - v_x, 
	a_{21} = \frac{2(b C_r - a C_f)}{I_z v_x}, 
	a_{22} = -\frac{2(a^2 C_f + b^2 C_r)}{I_z v_x}, 
	b_2 = \frac{1}{I_z},  
	e_1 = \frac{2 C_f}{m}, 
	e_2 = \frac{2 a C_f}{I_z}$.

Here, \(C_f\) and \(C_r\) are the front and rear tire cornering stiffness coefficients, \(a\) and \(b\) are the distances from the vehicle's center of gravity (CG) to the front and rear axles, respectively, \(m\) is the vehicle's total mass, \(I_z\) is the moment of inertia about the vertical axis, and \(v_x\) denotes the longitudinal velocity.

In practical applications, direct measurement of the vehicle's lateral velocity \(v_y\) is not available. 
Instead, the vehicle's sensors (e.g., Inertial Measurement Units) provide access to measurable quantities such as yaw rate, lateral or longitudinal acceleration. 
Accordingly, we consider four distinct output measurement cases.

\subsubsection{Case~1: Yaw rate as output}
In the first case, the vehicle's yaw rate \(r\) is assumed to be directly measurable. 
Thus, the output equation is expressed as
\begin{equation}
	y_1 = r,
\end{equation}
which yields the following output matrix:
\begin{equation}
	C_1 = 
	\begin{bmatrix}
		0 & 1
	\end{bmatrix}.
	\label{eq:C1}
\end{equation}

\subsubsection{Case~2: Lateral acceleration as output}
The second case considers the lateral acceleration \(a_y\) as the measurable quantity. 
From the lateral dynamics model \eqref{eq:lateral_dynamics}, the lateral acceleration can be expressed as
\begin{equation}
	a_y = \dot{v}_y + v_x r.
\end{equation}
Substituting from \eqref{eq:lateral_dynamics}, we have
\begin{equation}
	a_y = a_{11} v_y + (a_{12} + v_x) r + e_1 \delta,
\end{equation}
which leads to the output equation
\begin{equation}
	y_2 = 
	\begin{bmatrix}
		a_{11} & a_{12} + v_x
	\end{bmatrix} x + e_1 \delta.
\end{equation}
Hence, the corresponding output matrix is
\begin{equation}
	C_2 = 
	\begin{bmatrix}
		a_{11} & a_{12} + v_x
	\end{bmatrix}.
	\label{eq:C2}
\end{equation}

\subsubsection{Case~3: Combined yaw rate and lateral acceleration as outputs (Combined Linear Outputs)}
In the third case, both the yaw rate \(r\) and the lateral acceleration \(a_y\) are considered as outputs. 
The combined output vector is defined as
\begin{equation}
	y_3 = 
	\begin{bmatrix}
		r \\ a_y
	\end{bmatrix}
	=
	\begin{bmatrix}
		C_1 \\ C_2
	\end{bmatrix} x + 
	\begin{bmatrix}
		0 \\ e_1
	\end{bmatrix} \delta.
\end{equation}
Therefore, the overall output matrix for this case is given by
\begin{equation}
	C_3 =
	\begin{bmatrix}
		0 & 1 \\
		a_{11} & a_{12} + v_x
	\end{bmatrix}.
	\label{eq:C3}
\end{equation}

\subsubsection{Case 4: Longitudinal acceleration as output}
In this case, the longitudinal acceleration, obtained by IMU measurements is considered as the output of the system:
\begin{align}
	\label{eq:outputNonlinear}
    y(t)=\dot{v}_x-rv_y
\end{align}
Under constant speed assumptions, $\dot{v}_x=0$. We have the following:
\begin{align}
    y(t)=-rv_y
\end{align}
In this case, the output is a nonlinear function of the states, given by the product of $r$ and $v_y$.

Note that a 3-axis IMU can provide $r$ (yaw rate), $a_y$ (lateral acceleration), and $a_x$ (longitudinal acceleration) which will also be discuss subsequently.

\subsection{Resources and Objectives for Attacks on Vehicle Lateral Dynamics}
We characterize the attacker’s capabilities along three dimensions: system knowledge, disclosure, and disruption resources.

\textbf{System knowledge:} access to the plant and controller/observer models and gains (e.g., $A, B, C_i$ and gains), enabling model-based attack synthesis such as ZDA.

\textbf{Disclosure resources:} ability to read sensor and actuator data (e.g., IMU signals, steering angle, CAN frames), allowing recording and identification of system behavior.

\textbf{Disruption resources:} ability to inject, modify, or suppress signals in actuator or sensor channels (e.g., steering commands, CAN messages), enabling direct manipulation of the control loop.

\begin{figure}[t]
	\centering
	\includegraphics[width=0.9\linewidth]{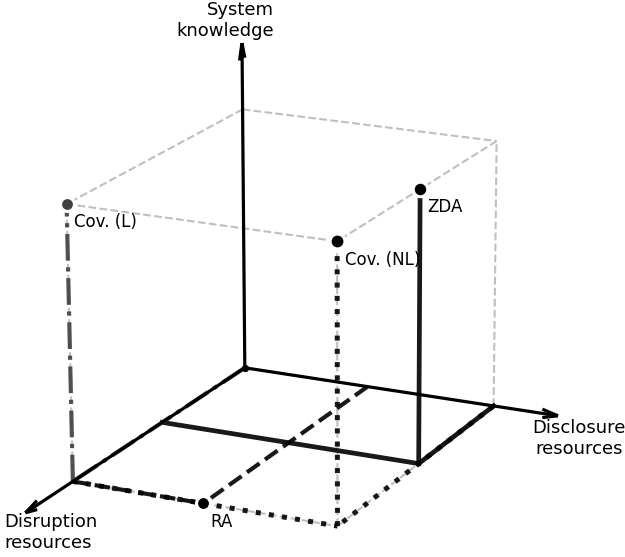}
	\caption{The stealthy attacks in the 3D resources space where RA, denotes replay attack, ZDA denotes zero dynamics attack, Cov. (L) denotes covert attacks with linear output ($r$ and $a_y$), and Cov. (NL) denotes covert attacks with nonlinear output ($a_x$).}
	\label{fig:Attack3d}
\end{figure}

Given the resources defined above, Fig. \ref{fig:Attack3d} illustrates the attack–resource space and positions the considered attacks according to their required resources.

Furthermore, the attacker's objectives can be summarized as (i) \emph{effectiveness}: to alter the vehicle's true physical state and cause degradation, and (ii) \emph{stealthiness}: to remain undetected by the monitoring mechanism, usually through ensuring that the output that is received at the controller is similar to the output in normal, attack-free scenarios.

In order to define stealthiness mathematically, let us denote the state and output in
normal, attack-free scenarios as $x^n(t) = [v_y^n(t),\, r^n(t)]^\top$ and
$y^n(t)$, respectively. Therefore, in normal, attack-free scenarios, the evolution of the system is as follows:
\begin{align}
	\label{eq:StealthinessPreparation}
	\begin{split}
		\dot{x}^n(t)=Ax^n(t)+Bu(t)\\
		y^n(t)=Cx^n(t)
	\end{split}
\end{align}
\textbf{Definition 1}: An attack signal is stealthy if the output received at the controller under attack, denoted as $y^*(t)$, is the same as the output in normal, attack-free scenarios, i.e.,
\begin{align}
	\label{eq:Stealthiness Definition}
	y^*(t)=y^n(t)
\end{align}

\section{Zero Dynamics Attacks Against Vehicle Lateral Dynamics}
\label{sec:zero_dynamics}

\begin{figure}[t]
    \centering
    \includegraphics[width=0.95\linewidth]{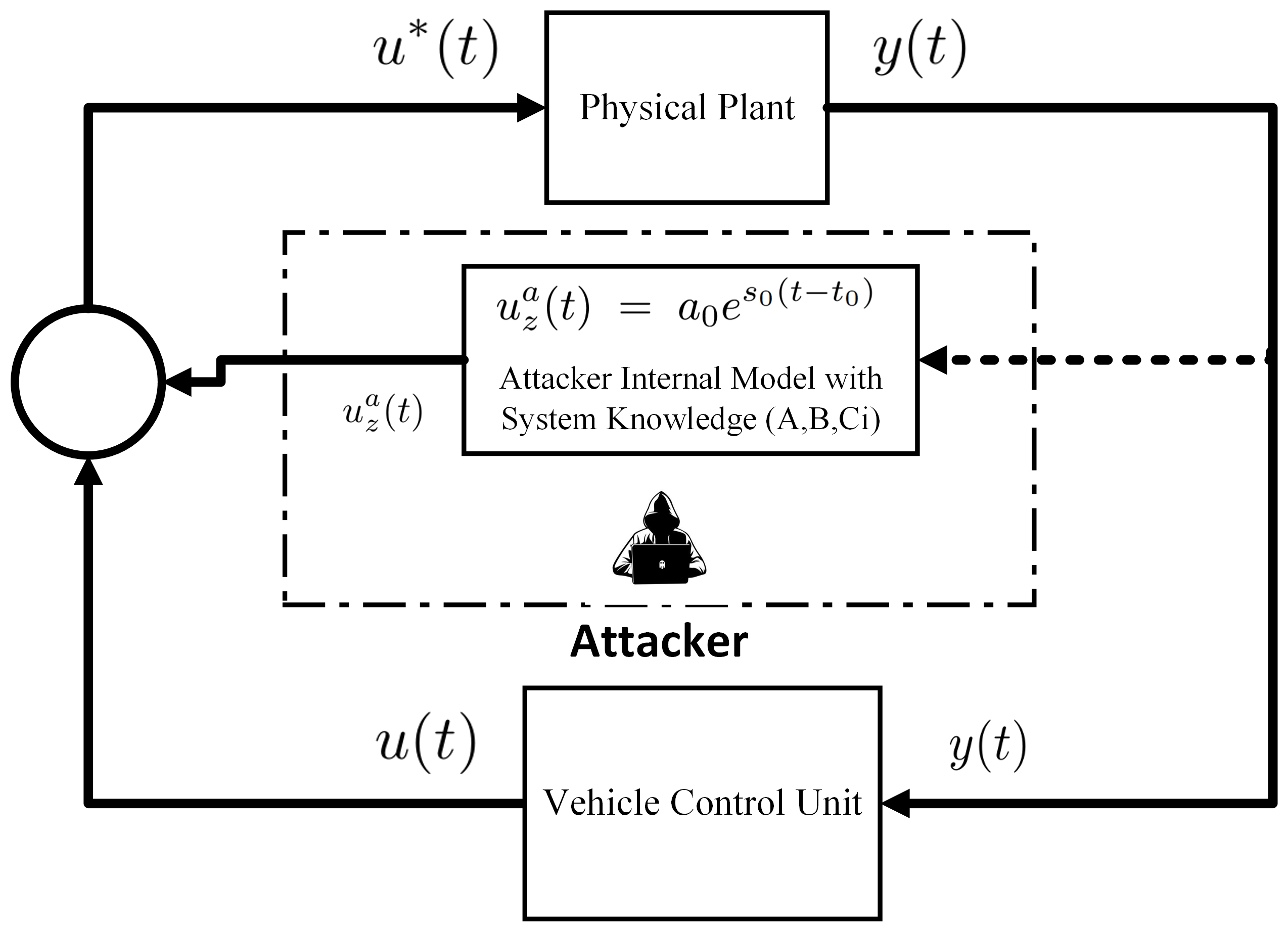}
    \caption{Zero Dynamics Attack (ZDA). the attacker computes the attack signal $u_{z}^{a}(t)$ and injects it into the actuator channels. Note that the main architecture of ZDA is similar in both linear and nonlinear outputs. The dashed line indicates the possibility of connection in the case where the states are not on the zero manifold.}
    \label{fig:ZDA}
\end{figure}

Zero Dynamics Attacks (ZDA) exploit invariant zeros to drive internal states without affecting the outputs. The architecture of ZDA is demonstrated in Fig. \ref{fig:ZDA}. 

\subsection{Linear Outputs ($r$ and $a_y$)}
Consider the linear time-invariant lateral dynamics (\ref{eq:lateral_dynamics}) where we consider only the yaw moment as the input:
\begin{align}
	\begin{split}
	\dot{x}(t) &= A x(t) + B M_z(t)\\
	y_i(t) &= C_i x(t)
	\end{split}
	\label{eq:lat_output}
\end{align}

Define the Rosenbrock (system) matrix
\[
P(s) \;=\; \begin{bmatrix} sI - A & -B \\ C_i & 0 \end{bmatrix}.
\]
An \emph{invariant zero} \(s_0\in\mathbb{C}\) is any complex value that renders \(P(s_0)\) rank deficient. Furthermore, there exists a nonzero vector \(z_0=(x_0,a_0)^\top\), \(x_0\in\mathbb{C}^n\), \(a_0\in\mathbb{C}^m\), such that
\begin{align}
	\begin{split}
	\label{eq:ZDAdefinition}
	(s_0 I - A) x_0 - B a_0 &= 0,\\
	C_i x_0 &= 0.
\end{split}
\end{align}
where $a_0$ is the attacker's initial condition that satisfies (\ref{eq:ZDAdefinition}). When such \((s_0,x_0,a_0)\) exist, the attack signal
\begin{equation}
	\label{eq:attack_input}
	u_{z}^{a}(t) \;=\;z_0e^{s_0 (t-t_0)},\quad t\ge t_0, 
\end{equation}
with $u_{z}^{a}(t)$ is the attack signal sent to the actuators. Then, this attack signal will result in:
\[
y_i(t) \;=\; C_ix(t) \equiv 0 \quad \text{for } t\ge t_0
\]

The following preposition from \cite{Shaaban2025CyberPhysicalSecurity} provides the zero dynamics attack signal for the different linear output cases.

The following proposition, adapted from Propositions~1-3 in \cite{Shaaban2025CyberPhysicalSecurity}, characterizes the zero-dynamics attack signals for different linear output selections.

\textbf{Proposition 1} \cite{Shaaban2025CyberPhysicalSecurity}:
Consider the linear state-space model (\ref{eq:lat_output}). The invariant zeros and
corresponding zero dynamics depend on the selected output as follows:
\begin{enumerate}
	\item \textit{Yaw rate output ($y=r$):}  
	The system has a single invariant zero
	$s_0=a_{11}$ and
	the corresponding zero dynamics attack input
$
	u_z^{a}=-\frac{a_{21}}{b_2}v_y
$
	excites this zero, resulting in a stable zero dynamics
$
	\dot v_y = a_{11} v_y.
$
	
	\item \textit{Lateral acceleration output ($y=a_y$):}  
	The system has the invariant zero
$
	s_0=\frac{a_{11}v_x}{a_{12}+v_x}.
$
	and the corresponding zero dynamics attack input
\begin{align*}
	u_z^{a}&=-\frac{1}{b_2(a_{12}+v_x)}
	(
	(a_{11}^2+a_{21}(a_{12}+v_x))v_y
	\\&+(a_{11}a_{12}+(a_{12}+v_x)a_{22})r
	)
\end{align*}
	excites this zero, yielding the zero dynamics
	\[
	\dot v_y = s_0 v_y, \qquad
	\dot r = s_0 r
	\]
	These zero dynamics are stable if and only if
	\[
	aC_f-bC_r<0.
	\]
	
	\item \textit{Combined yaw rate and lateral acceleration output
		($y=[r \; a_y]^\top$):}  
	The system has no invariant zeros. \hfill $\blacksquare$
\end{enumerate}

\textbf{Remark 1} (Non-Minimum-Phase Zeros and Stability in Lateral Vehicle Dynamics):
	The sign of the coupling term $aC_f-bC_r$ determines whether the lateral
	dynamics, when lateral acceleration is selected as the output, exhibit
	a non-minimum-phase zero. However, the existence of a non-minimum-phase
	zero does not imply instability of the system dynamics.
	The same coupling term appears in the off-diagonal entries of the state
	matrix $A$ in (\ref{eq:lateral_dynamics}). Nevertheless, open-loop stability of $A$ and the
	presence of a non-minimum-phase zero are distinct properties. For the
	state matrix $A$ in (1), Hurwitz stability is equivalent to:
	\begin{align}
		\operatorname{tr}(A) < 0, \qquad \det(A) > 0 .
	\end{align}
	These conditions depend on all entries of $A$, not solely on the term
	$aC_f-bC_r$. Consequently, parameter regimes may exist where
	$aC_f-bC_r > 0$ produces a non-minimum-phase zero while $A$ remains
	Hurwitz. Such situations may arise, for example, when the rear cornering
	stiffness $C_r$ degrades due to tire wear or reduced friction while
	$C_f$ remains approximately unchanged.

\subsection{Nonlinear output ($a_x$)}
The following preposition investigates zero dynamics attack with longitudinal acceleration ($a_x$) as the output and its proof is provided in Appendix \ref{sec:Appendix1}.

\textbf{Proposition 2}:
	Consider the linear vehicle model (\ref{eq:lateral_dynamics}) with nonlinear output (\ref{eq:outputNonlinear}). The zero-output manifold is
	\[
	\mathcal{Z} = \mathcal{Z}_1 \cup \mathcal{Z}_2, \quad \mathcal{Z}_1 = \{r,v_y|r = 0\}, \quad \mathcal{Z}_2 = \{r,v_y|v_y = 0\}.
	\]
	The zero dynamics on each branch are as follows:
	\begin{enumerate}
		\item On $\mathcal{Z}_1$ (with $r= 0$), the system has relative degree one, and the unique input enforcing $y(t)\equiv 0$ is
\begin{equation}
	\label{eq:attackzdanonlinearoutput}
		u_{z}^{a}(t)=\bar{M}_1 e^{a_{11} t}
\end{equation}
where
$
\bar{M}_1 := -\frac{a_{21}}{b_2} v_y(0).
$
		\item On $\mathcal{Z}_2$ (with $v_y = 0$), the only zero-output trajectory is the equilibrium
		$
		v_y(t) \equiv 0, \, r(t) \equiv 0, \, M_z(t) \equiv 0.
		$ \hfill $\blacksquare$
	\end{enumerate}

The structure of the zero-output manifold in Proposition 1 has important
implications for the internal behavior of the system. In particular,
the branch $\mathcal{Z}_2$ (defined by $v_y = 0$) admits no nontrivial
zero dynamics, since the only trajectory that maintains $y(t) \equiv 0$
is the equilibrium $(v_y,r)=(0,0)$. As a result, no nontrivial trajectories exist on this branch that keep the output equal to zero.

On the branch $\mathcal{Z}_1$ (defined by $r = 0$), zero dynamics do
exist and are governed by the internal mode $e^{a_{11}t}$ appearing in
(\ref{eq:attackzdanonlinearoutput}). For typical vehicle parameters,
$a_{11}<0$, which implies that this internal mode is exponentially
stable. Consequently, any zero dynamics attack associated with this mode
decays over time and its impact will be limited.

\begin{algorithm}[t]
	\caption{Zero-Dynamics Attack (ZDA) Policy for Linear Outputs}
	\label{alg:zda_linear_policy}
	\KwIn{Plant model $(A,B,C_i)$ where $A \in \mathbb{R}^{n \times n}$, $B \in \mathbb{R}^{n \times m}$, $C_i \in \mathbb{R}^{p \times n}$; attack start time $t_0$}
	\KwOut{Actuator-channel attack signal $u_z^a(t) \in \mathbb{R}^m$ for $t \geq t_0$}
	
	\textbf{Step 1: Invariant zero computation}\\
	Construct the Rosenbrock system matrix:
	\[
	P(s) = 
	\begin{bmatrix}
		sI - A & -B \\
		C_i & 0
	\end{bmatrix} \in \mathbb{C}^{(n+p) \times (n+m)}
	\]
	Find $s_0 \in \mathbb{C}$ such that $\mathrm{rank}(P(s_0)) < n+m$ by:\\
	\quad (a) Computing eigenvalues of $A$\\
	\quad (b) Testing each eigenvalue $\lambda_i$: if $\mathrm{rank}(P(\lambda_i)) < n+m$, set $s_0 = \lambda_i$\\
	\textbf{If} no invariant zero exists \textbf{then} \Return{Attack not feasible}
	
	\textbf{Step 2: Zero-direction computation}\\
	Solve the following for $(x_0, a_0) \neq (0,0)$:
	\[
	\begin{bmatrix}
		s_0 I - A & -B \\
		C_i & 0
	\end{bmatrix}
	\begin{bmatrix}
		x_0 \\
		a_0
	\end{bmatrix}
	= 0
	\]
	\textbf{Verify:} $C_i x_0 = 0$ (zero-output condition)
	
	\textbf{Step 3: Attack signal construction}\\
	\For{$t \geq t_0$}{
		Compute $u_z^a(t) = a_0\, e^{s_0(t - t_0)}$
	}
	
	\textbf{Step 4: Actuator injection}\\
	Inject $u_z^a(t)$ into the plant actuator channel: $u^*(t) = u(t) + u_z^a(t)$
	
	\textbf{Result:} The output satisfies $y_i(t) \equiv 0$ for $t \geq t_0$ while internal states diverge if $\mathrm{Re}(s_0) > 0$
\end{algorithm}

\begin{algorithm}[t]
	\caption{Zero-Dynamics Attack (ZDA) Policy for Nonlinear Output}
	\label{alg:zda_nonlinear_policy}
	\KwIn{Plant model $(A,B)$ where $A \in \mathbb{R}^{2 \times 2}$, $B = [0, b_2]^\top$; 
		output map $h(x) = -r\,v_y$;
		attack start time $t_0$;
		initial state estimate $\hat{x}(t_0) = [\hat{v}_y(t_0),\,\hat{r}(t_0)]^\top$;
		tolerance $\epsilon > 0$ (default: $10^{-6}$)}
	\KwOut{Actuator-channel attack signal $u_z^a(t) \in \mathbb{R}$ for $t \geq t_0$}
	
	\textbf{Step 1: Zero-manifold initialization}\\
	Evaluate zero-output condition: $\mathcal{Z} = \mathcal{Z}_1 \cup \mathcal{Z}_2$\\
	where $\mathcal{Z}_1 = \{r,v_y|r = 0\}$ and $\mathcal{Z}_2 = \{r,v_y|v_y = 0\}$\\
	
	\textbf{If} $|\hat{r}(t_0)| < \epsilon$ and $|\hat{v}_y(t_0)| \geq \epsilon$ \textbf{then}\\
	\quad State is on branch $\mathcal{Z}_1$ $\rightarrow$ proceed to Step 2\\
	\textbf{Else If} $|\hat{v}_y(t_0)| < \epsilon$ \textbf{then}\\
	\quad State is on branch $\mathcal{Z}_2$ (equilibrium) $\rightarrow$ \Return{No nontrivial attack}\\
	\textbf{Else}\\
	\quad \tcp{State not on zero manifold - preparatory phase required}
	\quad find steering input to reach $\mathcal{Z}_1$, then proceed to Step 2\\
	
	\textbf{Step 2: Attack signal construction on $\mathcal{Z}_1$}\\
	Compute
	$
	\bar{M}_1 = -\frac{a_{21}}{b_2}\,\hat{v}_y(t_0)
	$\\
	\For{$t \geq t_0$}{
		$u_z^a(t) = \bar{M}_1\,e^{a_{11}(t-t_0)}$
	}
	
	\textbf{Step 3: Actuator injection}\\
	Inject into plant: $u^*(t) = u(t) + u_z^a(t)$\\
\end{algorithm}

Note that zero dynamics attack policy for linear and nonlinear outputs are provided in Algorithm \ref{alg:zda_linear_policy} and \ref{alg:zda_nonlinear_policy}, respectively.

\textit{Attack Resources:}  
ZDA requires system knowledge of $(A,B,C_i)$ (and $h(x)$ in the nonlinear case) to compute the zero dynamics, along with disruption capability on the actuator channel. In practice, disclosure access to $(u(t), y(t))$ may be required for state estimation or to drive the system to the zero manifold.

\textit{Attack Impact:}  
The impact of ZDA is governed by the system zeros and depends on the selected output. For yaw rate output, the system exhibits a minimum-phase zero, and the internal states remain bounded. For lateral acceleration output, the behavior is determined by the term $aC_f - bC_r$, with stable or unstable zero dynamics depending on its sign. When both yaw rate and lateral acceleration are measured, no zero dynamics exist. For the nonlinear output $y = -r v_y$, the internal dynamics are stable ($a_{11} < 0$), and any attack-induced deviation decays over time.

\textbf{Remark 2}: Note that while the work in \cite{Shaaban2025CyberPhysicalSecurity} investigated ZDA in vehicle lateral dynamics, we have considered other stealthy attacks as well (i.e., replay attack and covert attacks). We have also investigated the case of nonlinear output (longitudinal acceleration) and the effects of saturation on the attack models, while only the linear output is investigated in \cite{Shaaban2025CyberPhysicalSecurity}.

\section{Covert Attacks Against Vehicle Lateral Dynamics}
\label{sec:Covert}
Covert attacks coordinate actuator and sensor attacks to alter system behavior while sending the nominal the controller receives the nominal outputs.

\begin{algorithm}[t]
	\caption{Covert Attack Policy for Linear Outputs}
	\label{alg:covert_linear_policy}
	\KwIn{Plant model $(A,B,C_i)$ where $A \in \mathbb{R}^{n \times n}$, $B \in \mathbb{R}^{n \times m}$, $C_i \in \mathbb{R}^{p \times n}$;
		attack start time $t_0$;
		actuator attack signal $u_c^a(t) \in \mathbb{R}^m$ for $t \geq t_0$;
		possible disclosure access to $y_i(t)$ and $u(t)$}
	\KwOut{Actuator signal $u^*(t)$ injected into plant;
		spoofed sensor signal $y^*(t)$ delivered to controller}
	
	\textbf{Prerequisite: Attack signal design (optional)}\\
	\textbf{If} $u_c^a(t)$ not provided, design via the following:\\
	\quad Choose attacker gains $k^a \in \mathbb{R}^{m \times n}$, $l^a \in \mathbb{R}^{m \times p}$\\
	\quad Define integral state: $z^a(t) = \int_{t_0}^{t} (y_i(s) - r^a(s))\,ds$ for target $r^a(t)$\\
	\quad Set: $u_c^a(t) = k^a \hat{x}(t) + l^a z^a(t) - u(t)$\\
	\quad where $\hat{x}(t)$ is estimated from observing $(u(t), y_i(t))$\\
	
	\textbf{Step 1: Attacker internal model initialization}\\
	Initialize attack state: $x^a(t_0) = 0 \in \mathbb{R}^n$\\
	Run attacker's dynamics for $t \geq t_0$:
	\[
	\dot{x}^a(t) = A x^a(t) + B u_c^a(t), \quad x^a(t_0) = 0
	\]
	
	\textbf{Step 2: Coordinated dual-channel injection}\\
	\For{$t \geq t_0$}{
		Inject modified signal: $u^*(t) = u(t) + u_c^a(t)$\\
		
		Compute attack-induced deviation: $y_c^a(t) = -C_i x^a(t)$\\
		Inject spoofed measurement: $y^*(t) = y_i(t) + y_c^a(t)$\\
	}

\end{algorithm}

\subsection{Linear Output}
\begin{figure}[t]
    \centering
    \includegraphics[width=0.95\linewidth]{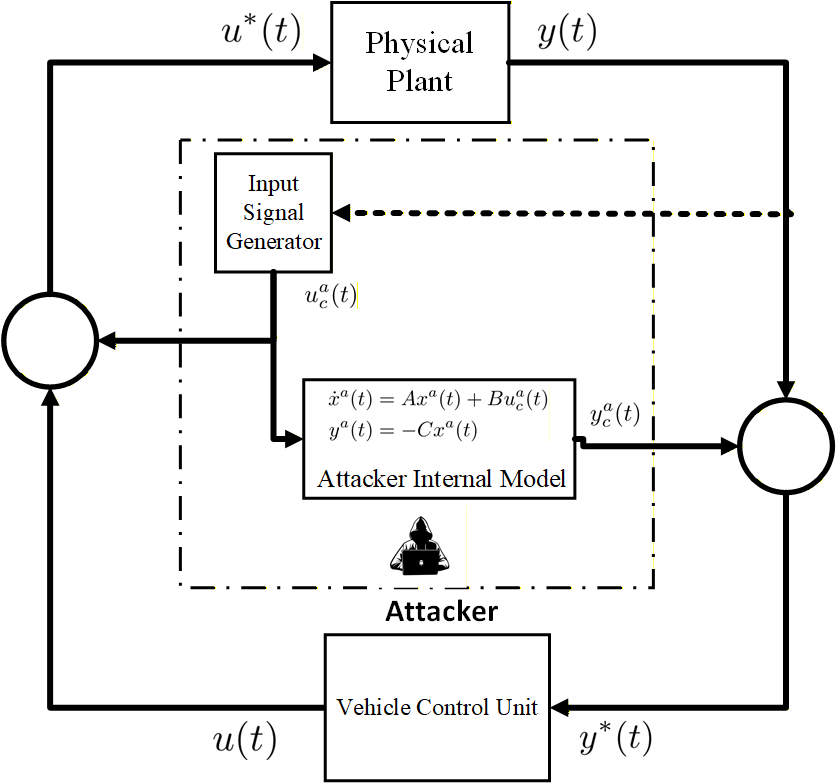}
    \caption{Covert attack architecture with linear output. Note that $x^{a}(0)=0$ and the dashed line represents possibility of connection depending on the design of the attacker's input signal $u_{c}^{a}(t)$.}
    \label{fig:CovertAttackLinear}
\end{figure}

Regardless of the design procedure for the attack signal $u_{c}^{a}$ (further information is provided in Remark 4), the sensor attack signal $y_{c}^{a}(t)$ is obtained through the following dynamics:
\begin{align}
	\label{eq:SensorAttack}
	\begin{split}
		\dot{x}_a=Ax_a+Bu_{c}^{a}\\
		y_{c}^{a}=-C_ix_a
	\end{split}
\end{align} 
where $x_a$ is the state of the attack with $x_a(0)=0$. it can be noted that in (\ref{eq:SensorAttack}), the sensor attack signal $y_{c}^{a}(t)$ is designed in coordination with the actuator attack signal $u_{c}^{a}(t)$. Figure \ref{fig:CovertAttackLinear} demonstrates covert attack's architecture when the outputs are linear.
One approach is to design $u_{c}^{a}$ to steer the system toward a desired trajectory $r_a(t)$ as follows:
\begin{align}
	u_{c}^{a}(t) &= k_a\hat{x}(t) + l_a z_a(t) - u(t) \label{eq:attack_design}\\
	z_a(t) &= \int_{t_0}^{t}(y(s) - r_a(s))ds \label{eq:z_a_equation}
\end{align}
where $k_a$ and $l_a$ are attacker-designed gains.
This construction requires disclosure access to $(u(t), y_i(t))$ and disruption capability on both actuator and sensor channels.

\subsection{Nonlinear Outputs}

\begin{algorithm}[t]
	\caption{Covert Attack Policy for Nonlinear Output ($a_x$)}
	\label{alg:covert_nonlinear_policy}
	\KwIn{Plant model $(A,B)$ where $A \in \mathbb{R}^{2 \times 2}$, $B = [0, b_2]^\top$; 
		output map $h(x) = -r\,v_y$;
		attack start time $t_0$;
		actuator attack signal $u_c^a(t)$;
		disclosure access to $u(t)$ and $y_i(t)$ for $t < t_0$}
	\KwOut{Actuator signal $u^*(t)$ injected into plant;
		spoofed sensor signal $y^*(t)$ delivered to controller}
	
	\textbf{Phase I: Pre-attack state estimation} ($t < t_0$)\\
	Use an estimator (e.g., Luenberger observer) on $(u(t), y_i(t))$ to obtain:
	\[
	\hat{x}^n(t) = [\hat{v}_y^n(t),\,\hat{r}^n(t)]^\top \quad \text{for } t \in [t_0 - T_{\text{obs}}, t_0]
	\]
	Continue estimation for $t \geq t_0$ using $u(t)$ and internal dynamics\\
	
	\textbf{Phase II: Attack execution} ($t \geq t_0$)\\
	
	\textbf{Step 1:} Initialize attack state: $x^a(t_0) = 0$\\
	Run attacker's dynamics for $t \geq t_0$:
	\[
	\dot{x}^a(t) = A x^a(t) + B u_c^a(t), \quad x^a(t) = [v_y^a(t),\, r^a(t)]^\top
	\]
	
	\textbf{Step 2:} Coordinated dual-channel injection:\\
	\For{$t \geq t_0$}{
		Inject: $u^*(t) = u(t) + u_c^a(t)$\\
		Compute: $y_c^a(t) = v_y^a(t)\,\hat{r}^n(t) + \hat{v}_y^n(t)\,r^a(t) + 2\,v_y^a(t)\,r^a(t)$\\
		Inject: $y^*(t) = y_i(t) + y_c^a(t)$\\
	}

\end{algorithm}

\begin{figure}[t]
	\centering
	\includegraphics[width=0.95\linewidth]{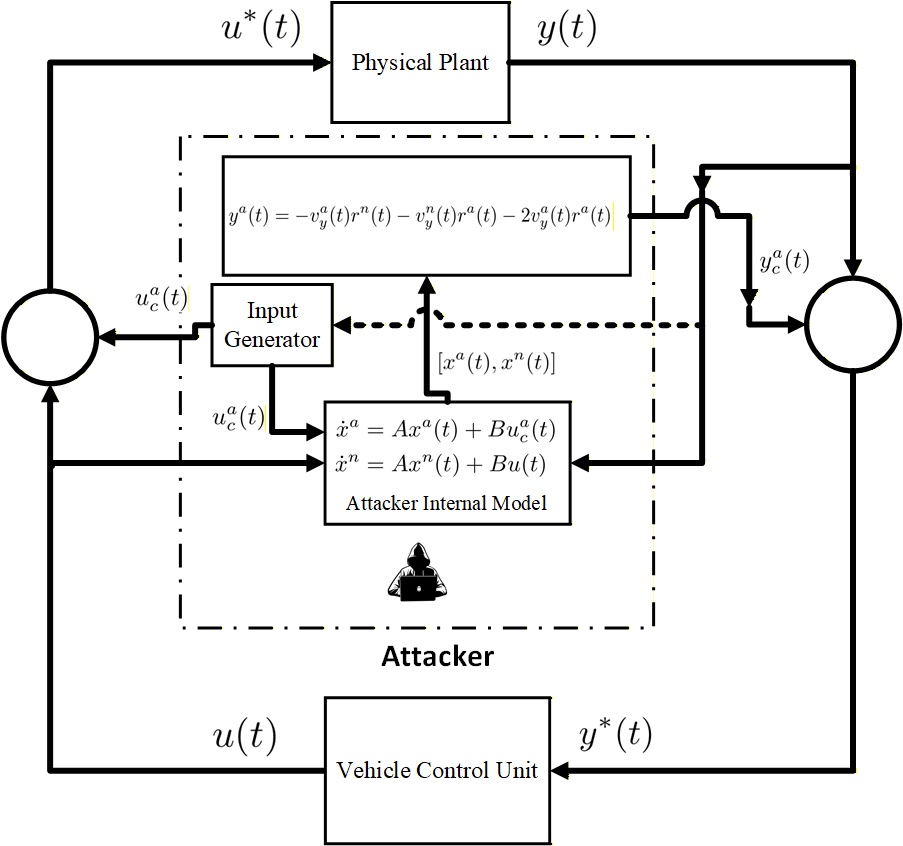}
	\caption{Covert attack when the output is nonlinear (longitudinal acceleration). Note that the internal model also includes an estimator to obtain the states of the physical plant (the first phase of covert). Furthermore, the dashed line indicates possible connection depending on the design of the attacker's input signal $u_{c}^{a}(t)$.}
	\label{fig:CovertAttackNonlinear}
\end{figure}
The overall architecture of covert attacks in case of nonlinear output is provided in Fig. \ref{fig:CovertAttackNonlinear}.
Given that the output is $y(t) = -r(t)v_y(t)$, we have $v_y(t) = [1,\,0]x(t) = F_1x(t)$
and $r(t) = [0,\,1]x(t) = F_2x(t)$, so that
\begin{align}
	y(t) = -F_1x(t)F_2x(t),
\end{align}
with $F_1 = [1,\,0]$ and $F_2 = [0,\,1]$. Under a covert attack, the attacker injects
$u^*(t) = u(t) + u_c^a(t)$ into the actuator channel, and maintains an internal
model
\begin{align}
	\dot{x}^a(t) = Ax^a(t) + Bu_c^a(t), \quad x^a(0) = 0,
\end{align}
with $x^a(t) = [v_y^a(t),\, r^a(t)]^\top$. The following preposition discusses the required attack signal on the output channel to ensure stealthiness of the attacker with its proof provided in Appendix \ref{sec:Appendix2}.

\textbf{Preposition 3}: Consider the lateral dynamics (\ref{eq:lateral_dynamics}) with the nonlinear output (\ref{eq:outputNonlinear}). For the covert attack to remain stealthy based on definition 1 (i.e., $y^*(t)=y^n(t)$), the attacker must inject the following output compensation signal
	$y_c^a(t)$ into the output channel:
	\begin{align}
		\label{eq:CovOutputFinal}
		y_c^a(t) = v_y^a(t)\,r^n(t) + v_y^n(t)\,r^a(t) + 2v_y^a(t)\,r^a(t)
	\end{align}\hfill $\blacksquare$

Note that the covert attack policy for linear and nonlinear outputs are provided in Algorithm \ref{alg:covert_linear_policy} and \ref{alg:covert_nonlinear_policy}, respectively.

\textit{Attack Resources:}  
Covert attacks require simultaneous access to actuator and sensor channels, enabling coordinated input injection and output manipulation. 
For linear outputs, knowledge of $(A,B,C_i)$ (or an approximate model) is required to compute the output attack signal $y_{c}^{a}$. Disclosure access to $(u(t), y_i(t))$ is only necessary if the attacker aims to enforce reference tracking.

For nonlinear outputs, stronger model knowledge is required, including both the state dynamics and output map. In this case, the attacker must estimate the nominal states $(r^n(t), v_y^n(t))$, typically via an observer, to compute the compensation signal in \eqref{eq:CovOutputFinal}.

\textit{Attack Impact:}  
The impact of covert attacks depends on the attacker's objective but is generally significant due to access to both actuator and sensor channels. As a result, the attacker can induce behaviors ranging from increased probability of lane deviations~\cite{farivar2021covert} to full tracking of a desired reference signal. In this sense, covert attacks effectively provide complete control authority over the lateral dynamics.

\section{Replay Attack Against Vehicle Lateral Dynamics}
\label{sec:replay}
A replay attack occurs if an adversary replaces the current measurement with a previously recorded one, i.e., $
y_i^*(t) = y_{i}(t-\tau)$ for some delay $\tau > 0$, where $y_{t-\tau}$ was recorded during an earlier normal, attack-free operation period. The attacker can also attack the input channel to cause degradation through $u^*(t)=u(t)+u_{r}^{a}(t)$. 

\begin{figure}[t]
	\centering
	\includegraphics[width=0.9\linewidth]{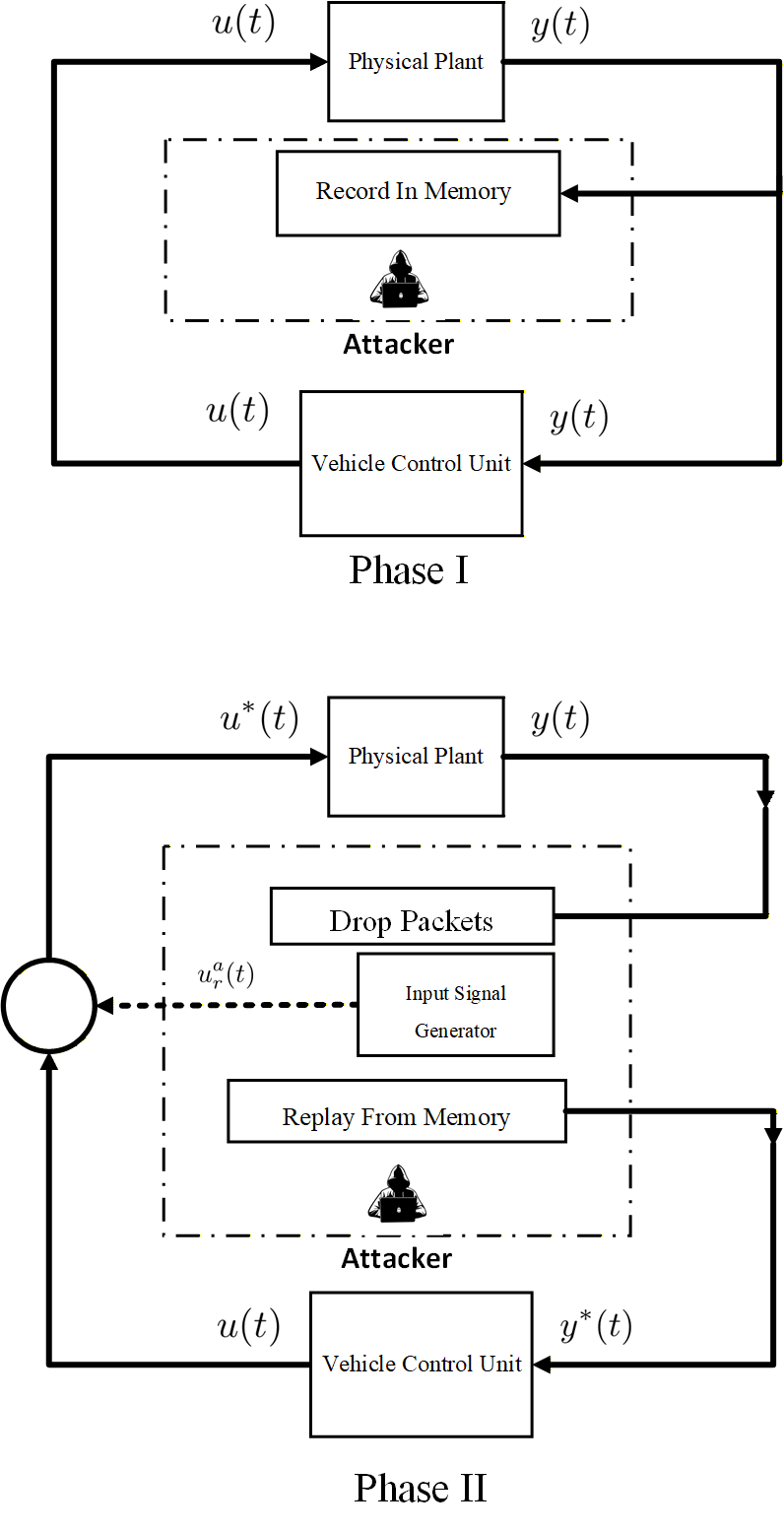}
	\caption{Replay Attack procedure with both linear and nonlinear outputs. The dashed line represents the possibility of actuator channel attacks.}
	\label{fig:Replay}
\end{figure}

\begin{algorithm}[t]
	\caption{Replay Attack Policy}
	\label{alg:replay_policy}
	\KwIn{Disclosure access to sensor output $y_i(t)$;
		recording window duration $\tau > 0$;
		recording end time $t_r$ (attack start: $t_0 = t_r$);
		steady-state tolerance $\delta > 0$;
		optional actuator injection signal $u_{r}^{a}(t)$}
	\KwOut{Spoofed output $y_{i}^{*}(t)$ delivered to controller;
		actuator signal $u^*(t)$ seen by plant (if actuator attack included)}
	
	\textbf{Phase I: Recording steady-state window} ($t \in [t_r - \tau, t_r]$)\\
	
	\textbf{Step 1:} Verify steady-state condition:\\
	\quad Check: $\|\dot{y}_i(t)\| < \delta$ for all $t \in [t_r - \tau, t_r]$\\
	\quad \textbf{If} not satisfied \textbf{then} select different recording window\\
	
	\textbf{Step 2:} Record sensor measurements:
	\[
	\mathcal{R} = \{y_i(t) : t \in [t_r - \tau,\, t_r]\}
	\]
	
	\textbf{Phase II: Replay and optional actuator injection} ($t \in [t_r, t_r + \tau]$)\\
	
	\For{$t \in [t_r,\, t_r + \tau]$}{
		Block sensor packet $y_i(t)$ from reaching controller\\
		Retrieve from buffer: $y_{i}^{*}(t) = y_i(t - \tau)$ \\
		Transmit $y_{i}^*(t)$ to controller\\
		
		\textbf{If} $u_r^a(t)$ provided \textbf{then}\\
		\quad Inject: $u^*(t) = u(t) + u_r^a(t)$\\
		\textbf{Else}\\
		\quad $u^*(t) = u(t)$ \tcp*{No actuator attack}
	}

\end{algorithm}

Note that the replay attack policy is the same for the case of linear and nonlinear outputs, and the policy is provided in Algorithm \ref{alg:replay_policy}.

\textit{Attack Resources:}  
Replay attacks do not require system model knowledge, placing them at the lowest level of system knowledge in Fig.~\ref{fig:Attack3d}. Instead, they rely on communication access: disclosure capability to record sensor measurements $y_i(t)$ during normal operation, and disruption capability to suppress and replay these measurements. To remain stealthy, the recorded data must correspond to a steady-state operating regime and if the attacker aims to alter the physical behavior, additional actuator-channel access is required to inject $u_r^{a}(t)$.

\textit{Attack Impact:}  
Replay attacks can induce significant performance degradation while remaining stealthy. Small discrepancies between the replayed and true measurements can cause the estimator and controller to converge to incorrect states, leading to drift and potential loss of lateral stability. Over time, these errors accumulate and may result in large yaw deviations or lane departures. While replay attacks can be combined with actuator injection to amplify their effect, the lack of system knowledge limits the attacker’s ability to enforce a specific trajectory.

\section{Effects of Actuator and Tire Saturation}
\subsection{Effects of Saturation on Attack Stealthiness and Impact}

Actuator and tire saturation introduce nonlinearities that affect both the stealthiness and effectiveness of attacks on vehicle lateral dynamics. The impact, however, differs significantly across attack classes.

For ZDA, actuator saturation clips the injected signal, disrupting the zero-dynamics structure and removing stealthiness. At the same time, it bounds the internal state evolution, preventing divergence and reducing attack impact.

In contrast, replay attacks are largely insensitive to saturation, as they rely on recorded sensor measurements rather than precise actuator shaping. Although actuator limits may restrict the extent of physical degradation when actuator injection is present, the replayed outputs remain consistent with steady-state behavior, thereby preserving stealthiness.

Covert attacks exhibit a different behavior due to the attacker’s access to both actuator and sensor channels. This enables coordinated input injection and output manipulation, so actuator saturation does not fundamentally limit the attack: inputs can be designed within saturation bounds while their effects are masked at the output. However, if the attacker lacks knowledge of the saturation limits, this coordination may fail, leading to detectability. An exception is mirror attacks~\cite{mikhaylenko2024robust}, where the output is directly replaced and stealthiness is preserved even under saturation. 
Finally, the temporal structure of covert attacks also plays an important role. Low-magnitude, long-duration inputs can accumulate significant state deviations, while saturation may delay recovery after the attack ends, resulting in more persistent effects.

\subsection{Saturation-Induced Loss of Control Authority and Model Validity}
Another degradation pathway arises when an attacker drives the system into a saturation regime, activating anti-windup mechanisms in the controller. While these mechanisms prevent integrator accumulation, their activation reduces control authority and responsiveness. An attacker can exploit this effect to induce prolonged transients or delayed recovery, even without sustained large inputs. More generally, under saturation, additional control effort no longer yields proportional lateral force or yaw moment, leading to a loss of control authority.

Saturation also affects the validity of the system model. In practice, tire saturation often occurs before steering actuator limits are reached, causing a collapse in effective cornering stiffness. This invalidates the linear lateral dynamics model commonly used for control and attack design. In this regime, both linear controllers and attack strategies no longer reflect the true system behavior, and inputs computed under linear assumptions may produce degraded responses. Consequently, both control recovery and attack detection become more challenging due to model mismatch.

\section{Simulation Case Studies}
 
\begin{figure}[t]
    \centering
    \includegraphics[width=\linewidth]{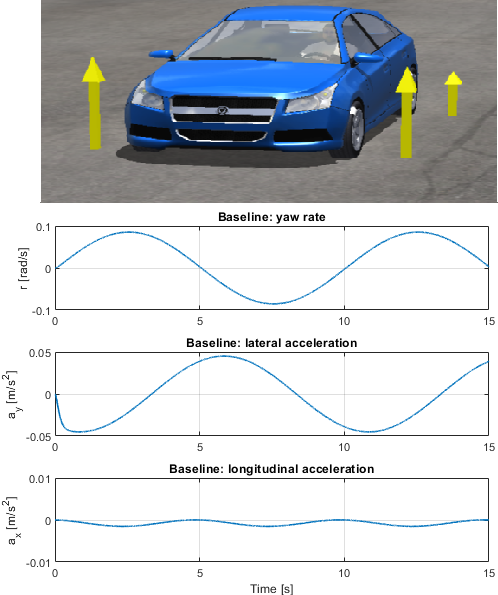}
    \caption{Outputs of a class-C 2017 hatchback vehicle in normal, attack-free scenarios in CarSim when the vehicle is moving with sinusoidal input.}
    \label{fig:Normal}
\end{figure}

\begin{table}[h!]
	\centering
	\caption{Vehicle parameters}
	\label{tab:vehicle_params_linear}
	\begin{tabular}{l c c}
		\hline
		\textbf{Parameter} & \textbf{Symbol} & \textbf{Value} \\
		\hline
		Vehicle mass & $m$ & $1412~\mathrm{kg}$ \\
		
		Yaw moment of inertia & $I_z$ & $1536.7~\mathrm{kg\,m^2}$ \\
		
		Distance CG to front axle & $a$ & $1.015~\mathrm{m}$ \\
		
		Distance CG to rear axle & $b$ & $1.895~\mathrm{m}$  \\
		
		Wheelbase & $L$ & $2.910~\mathrm{m}$  \\
		
		Longitudinal speed (linearization point) & $V_x$ & $16.67~\mathrm{m/s}$  \\
		
		Front cornering stiffness & $C_f$ & $58.4~\mathrm{kN/rad}$  \\
		
		Rear cornering stiffness & $C_r$ & $40.4~\mathrm{kN/rad}$  \\
		\hline
	\end{tabular}
\end{table}
In order to verify our analysis and the different attacks discussed in this paper, we investigate our results through CarSim/Simulink co-simulation. We have considered a class-C 2017 hatchback vehicle with Table \ref{tab:vehicle_params_linear} providing some of the parameters of this vehicle. In this section, we will consider the three attack types discussed in previous sections. The baseline scenario is the car following sinusoidal steering pattern with Fig. \ref{fig:Normal} demonstrating the outputs of the lateral dynamics when the system is not subject to any attacks (Normal situation).
\subsection{Replay Attack}
\begin{figure}[t]
	\centering
	\includegraphics[width=\linewidth]{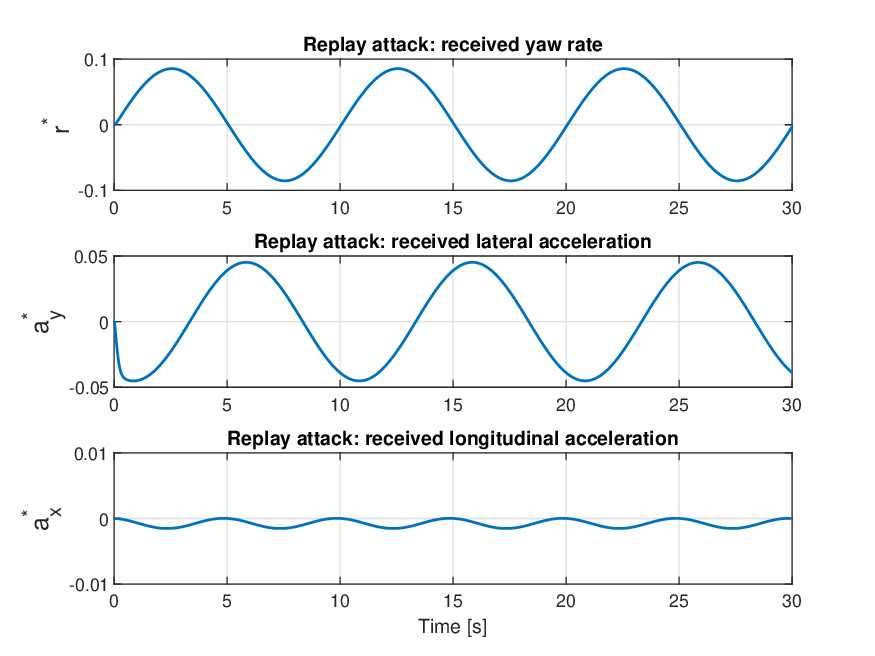}
	\caption{Replayed Outputs received at the control unit. As demonstrated, the replayed outputs show no signs of anomalies.}
	\label{fig:Replayed1}
\end{figure}

\begin{figure}[t]
	\centering
	\includegraphics[width=\linewidth]{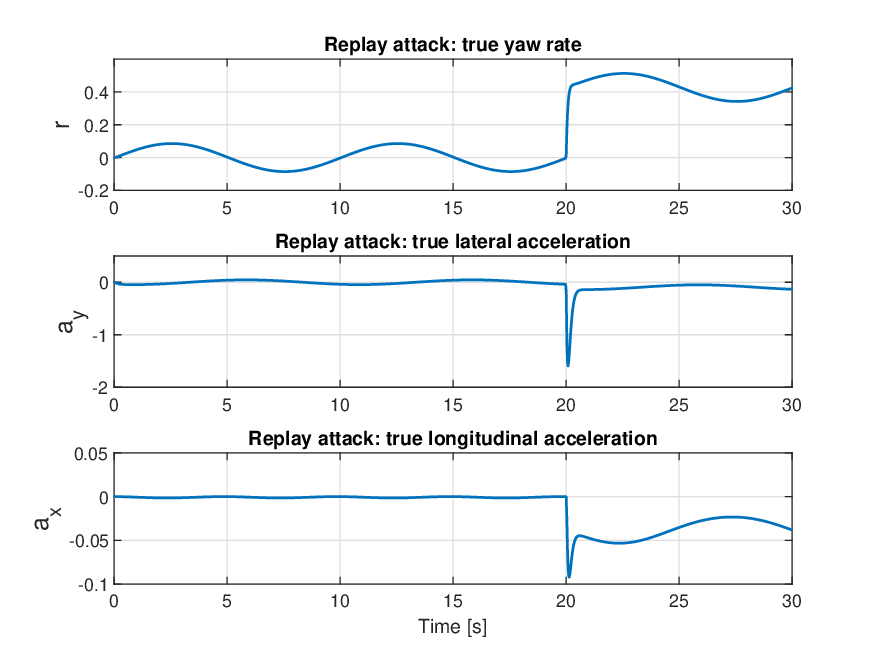}
	\caption{True outputs under replay attack, demonstrating the capabilities of replay attacks in degrading the performance of the system.}
	\label{fig:Replayed2}
\end{figure}

First, we consider replay attacks on the outputs of the system. The attacker records the outputs for 10 second from $t=10$ and then repeats the recorded value while injecting a FDI attack on the steering angle with $u_{r}^{a}(t>20)=5$. Figure \ref{fig:Replayed1} demonstrates the replayed outputs received at the control unit while Fig. \ref{fig:Replayed2} shows the true outputs of the system when the system is subject to replay attacks. As demonstrated in these two figures, the attacker can degrade the system performance without any requirements on the system knowledge.

\subsection{Zero Dynamics Attack}
\begin{figure}[t]
	\centering
	\includegraphics[width=\linewidth]{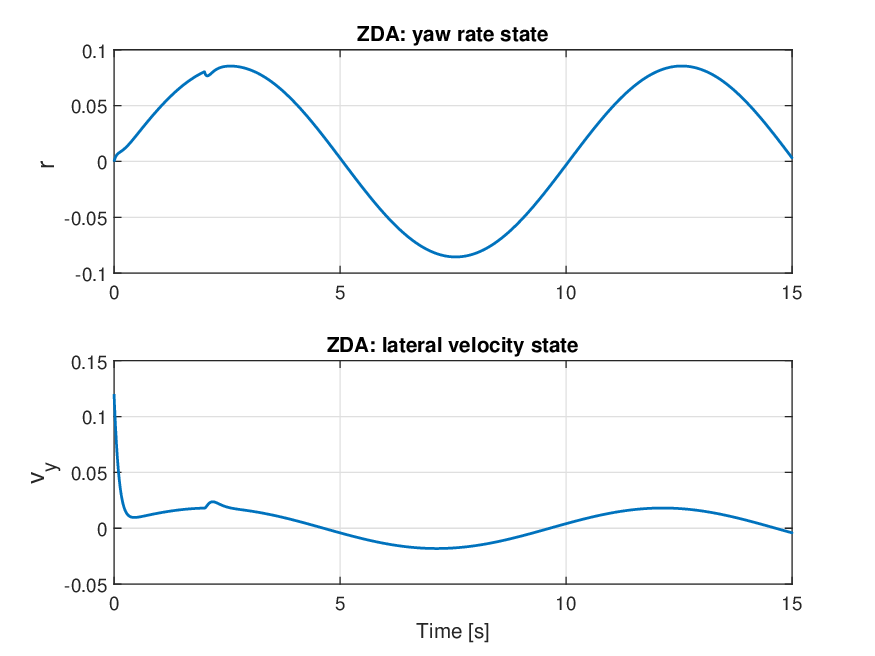}
	\caption{ States of the lateral dynamics subject to ZDA. Since the system has stable zeros, the ZDA will not cause instability in the system.}
	\label{fig:ZDA2}
\end{figure}

The corresponding zero of the considered vehicle (class-C hatchback) for the case of yaw rate ($r$) as the output is $s_0=-4.9218$ and for the case of lateral acceleration ($a_y$) as the output is $s_0=-3.4941$. Therefore, the corresponding zero dynamics attack signals will converge to zero. For the case of longitudinal acceleration, given that in equation (\ref{eq:attackzdanonlinearoutput}), the signal is of the form $\bar{M}_1 e^{a_{11} t} $, since $a_{11}=-4.1983$, the attack signal will converge to zero as well. Figure \ref{fig:ZDA2} demonstrates the states of the system subject to ZDA, showing that the ZDA signal will not cause instability in the system.

\subsection{Covert Attacks}
\begin{figure}[t]
	\centering
	\includegraphics[width=\linewidth]{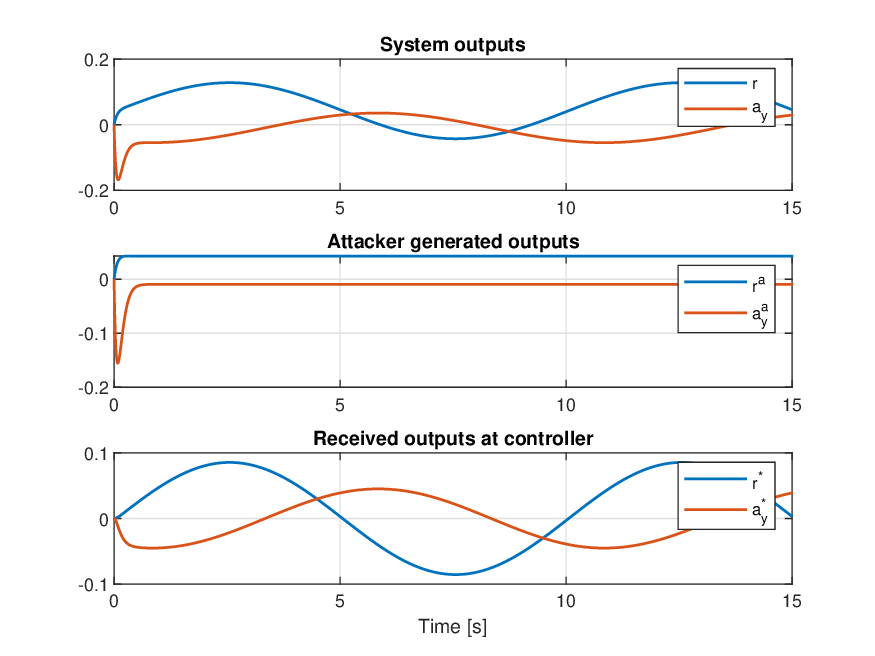}
	\caption{The outputs of the system (top figure), generated outputs by the attacker (middle figure), and the received outputs at the controller (bottom figure). As demonstrated, the attacker is capable of removing the effects of $u_{c}^{a}$ by injecting $y_{c}^{a}$ almost completely. However, since the linear model cannot capture the complexities of the actual system model, the received outputs will be slightly different than the outputs in normal, attack-free scenarios.}
	\label{fig:Covert1}
\end{figure}

In this section, we consider the case of covert attacks on the steering angle with $u_{c}^{a}(t>=0)=1$. For the case of linear outputs (yaw rate and lateral acceleration), the attacker computes the output signal $y_{c}^{a}(t)$ based on equation (\ref{eq:SensorAttack}) and then injects it on the output channel. Figure \ref{fig:Covert1} demonstrates the system outputs under covert attack (top figure), the outputs generated by the attacker (middle figure), and the received outputs at the control unit (bottom figure). As can be seen, the attacker can generate the linear outputs that are close to the outputs of the system. Note that there exist a small change in the received output at the controller side in comparison to the normal, attack-free scenarios. This shows both the capabilities and limitations of the linear model in capturing the dynamics of the system.

\begin{figure}[t]
	\centering
	\includegraphics[width=\linewidth]{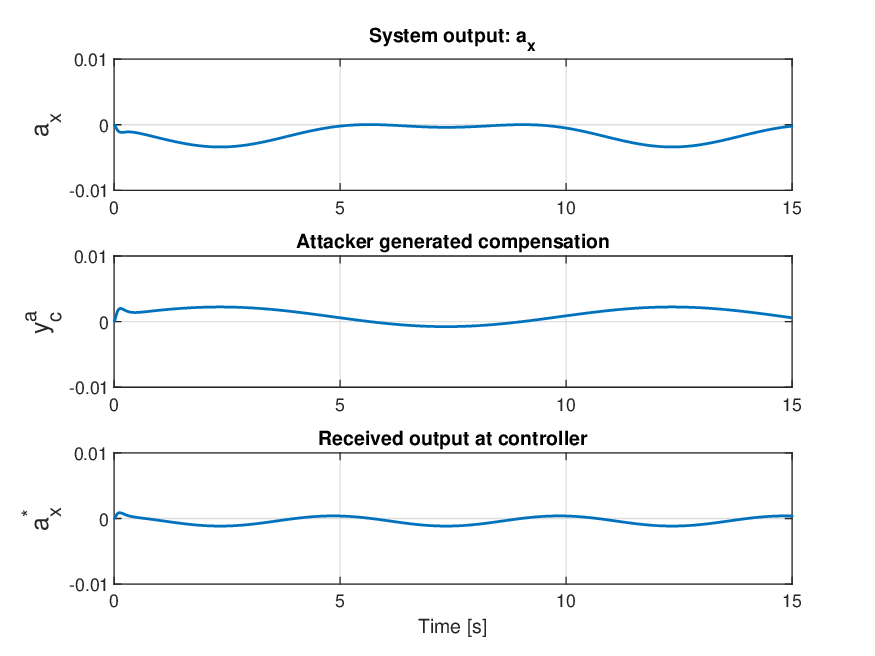}
	\caption{Nonlinear output (i.e., longitudinal acceleration $a_x$) with system output (top figure), generated signal by the attacker (middle figure), and the received output at the controller (bottom figure). The effects of the covert attack are slightly visible in the system output received at controller.}
	\label{fig:Covert2}
\end{figure}

For the case of nonlinear output (longitudinal acceleration), Fig. \ref{fig:Covert2} demonstrated the system output (top figure), the generated output by the attacker (middle figure) and the received output at the controller side (bottom figure). It can be seen that the attack signal $u_{c}^{a}$ influences the longitudinal acceleration much less than the linear outputs with only a slight difference between the outputs in normal attack-free scenarios and the outputs under covert attacks.

\section{Conclusion \& Future Work}

This paper examined stealthy attacks on vehicle lateral dynamics from a system-theoretic perspective, focusing on replay attacks, ZDA, and covert attacks within a two-degree-of-freedom bicycle model. By analyzing multiple output selections, such as yaw rate, lateral acceleration, and nonlinear longitudinal acceleration, we characterized each attack in terms of required system knowledge, communication access, and the impact on the system. Furthermore, effects of saturation are also investigated on the attack impact and stealthiness.

Future research will extend this analysis to other stealthy attacks such as Pole Dynamics Attacks. Furthermore, developing detection and mitigation strategies by taking into account the resources and impact of each type of attack is an important research direction. One can investigate detection approaches such as Watermarking \cite{ghamarilangroudi2025replay}, coding schemes \cite{liu2025matrix}, and auxiliary system approaches \cite{eslami2024event} in the context of vehicle lateral dynamics and analyze the feasibility and performance of these approaches against stealthy attacks. Finally, as agentic AI systems are increasingly deployed across vehicle subsystems, including infotainment and control, their application to attack detection and isolation presents promising opportunities given demonstrated performance across diverse attack types. However, their adoption also introduces novel threat surfaces, including memory poisoning, prompt injection, and hallucination-induced actions \cite{eslami2025security}, which will be systematically characterized, detected, and mitigated in our future works.

\appendices
\section{Proof of Proposition 2}
\label{sec:Appendix1}
We consider the nonlinear output
\begin{equation}
	y = h(x) = -r\,v_y,
	\label{eq:nonlinear_output_zero_dyn}
\end{equation}
which couples the lateral velocity and yaw rate. This defines a nonlinear SISO system of the form
\[
\dot{x} = f(x) + g(x) M_z, \qquad y = h(x),
\]
with
\[
f(x) = A x, \qquad g(x) = B.
\]
In order to find the zero-output manifold, we have:
\begin{equation}
	y = h(x) = -r\,v_y = 0.
	\label{eq:zero_manifold_condition_zero_dyn}
\end{equation}
Geometrically, this defines the zero-output manifold
\begin{align*}
	\mathcal{Z}
	&= \{ x \in \mathbb{R}^2 : r\,v_y = 0 \}
	\\&= \{ (v_y, r) : r = 0 \} \cup \{ (v_y, r) : v_y = 0 \}.
	\label{eq:zero_manifold_zero_dyn}
\end{align*}
Thus, \(\mathcal{Z}\) is the union of two lines in the \((v_y,r)\)-plane:
\[
\text{branch } \mathcal{Z}_1: r = 0, \quad \text{branch } \mathcal{Z}_2: v_y = 0.
\]

Zero dynamics describe the internal motion of the system when trajectories are constrained to remain on \(\mathcal{Z}\), i.e., when \(y(t)\equiv 0\) for all \(t\). 
The gradient of the output \eqref{eq:nonlinear_output_zero_dyn} is
\begin{equation}
	\nabla h(x)
	=
	-\begin{bmatrix}
		\partial h / \partial v_y \\[1pt]
		\partial h / \partial r
	\end{bmatrix}
	=
	-\begin{bmatrix}
		r \\[1pt] v_y
	\end{bmatrix}.
	\label{eq:grad_h_zero_dyn}
\end{equation}
The Lie derivatives along \(f\) and \(g\) are
\begin{align*}
	L_f h(x)
	&= \nabla h(x)^{\top} f(x)
	= -\begin{bmatrix} r & v_y \end{bmatrix} A x, \label{eq:Lf_general_zero_dyn}\\[2pt]
	L_g h(x)
	&= \nabla h(x)^{\top} g(x)
	= -\begin{bmatrix} r & v_y \end{bmatrix} B
	= -r \cdot 0 + v_y b_2
	\\&= -b_2 v_y.
\end{align*}
Using the entries of \(A\), the Lie derivative can be written explicitly as
\begin{align}
	L_f h(x)
	&= -\begin{bmatrix} r & v_y \end{bmatrix}
	\begin{bmatrix}
		a_{11} & a_{12} \\
		a_{21} & a_{22}
	\end{bmatrix}
	\begin{bmatrix}
		v_y \\ r
	\end{bmatrix}\\\nonumber
	&= -(a_{11} r v_y + a_{12} r^2 + a_{21} v_y^2 + a_{22} r v_y)
	\label{eq:Lf_expanded_zero_dyn}
\end{align}
The first derivative of the output is therefore
\begin{align}
	\dot{y}
	&= L_f h(x) + L_g h(x) M_z
	\\&= -(a_{11} r v_y + a_{12} r^2 + a_{21} v_y^2 + a_{22} r v_y + b_2 v_y M_z).
	\label{eq:y_dot_general_zero_dyn}
\end{align}

For points where \(v_y \neq 0\), we have \(L_g h(x) = b_2 v_y \neq 0\), and the system has relative degree one with respect to \(y\). In that region, enforcing \(\dot{y}=0\) allows us to solve for the input as a state-dependent feedback
\begin{equation}
	M_z^{\star}(x)
	= \frac{L_f h(x)}{L_g h(x)}
	= \frac{a_{11} r v_y + a_{12} r^2 + a_{21} v_y^2 + a_{22} r v_y}{b_2 v_y}
	\label{eq:M_star_general_zero_dyn}
\end{equation}
The zero dynamics are then the closed-loop dynamics
\begin{equation}
	\dot{x} = A x + B M_z^{\star}(x),
	\label{eq:closed_loop_zero_dyn_general}
\end{equation}
restricted to the zero manifold \(h(x)=0\). Because the manifold \(\mathcal{Z}\) is the union of \(\{r=0\}\) and \(\{v_y=0\}\), we analyze each branch separately.

\textbf{(i)} On the first branch ($r=0$) we impose
\begin{equation}
	r(t) \equiv 0, \qquad v_y(t) \neq 0,
	\label{eq:branch1_def_zero_dyn}
\end{equation}
which ensures \(y(t) = -r(t)v_y(t) = 0\) for all \(t\). Using the original state equations,
\begin{align}
	\dot{v}_y &= a_{11} v_y + a_{12} r, \label{eq:vy_dyn_zero_dyn}\\
	\dot{r}   &= a_{21} v_y + a_{22} r + b_2 M_z, \label{eq:r_dyn_zero_dyn}
\end{align}
the output derivative can be written via the product rule:
\begin{equation}
	\dot{y}
	= -\frac{d}{dt}(r v_y)
	= -r \dot{v}_y - v_y \dot{r}.
	\label{eq:y_dot_product_zero_dyn}
\end{equation}
Substituting \eqref{eq:vy_dyn_zero_dyn}–\eqref{eq:r_dyn_zero_dyn} and then enforcing \(r=0\) yields
\begin{align}
	\dot{y}\big|_{r=0}
	&= -r (a_{11} v_y + a_{12} r)
	- v_y (a_{21} v_y + a_{22} r + b_2 M_z)\Big|_{r=0} \nonumber\\[2pt]
	&= -v_y \big(a_{21} v_y + b_2 M_z\big).
	\label{eq:y_dot_branch1_zero_dyn}
\end{align}
Enforcing \(\dot{y}=0\) on this branch and assuming \(v_y \neq 0\) leads to
\begin{equation}
	a_{21} v_y^2 + b_2 v_y M_z = 0
	\quad \Rightarrow \quad
	M_z^{\star}(v_y, r=0)
	= - \frac{a_{21}}{b_2} v_y.
	\label{eq:M_star_branch1_zero_dyn}
\end{equation}
Substituting \(r=0\) and \eqref{eq:M_star_branch1_zero_dyn} into the state dynamics gives the zero dynamics on \(\mathcal{Z}_1\):
\begin{align}
	\dot{v}_y
	&= a_{11} v_y + a_{12} r
	= a_{11} v_y, \label{eq:vy_branch1_zero_dyn}\\[2pt]
	\dot{r}
	&= a_{21} v_y + a_{22} r + b_2 M_z^{\star}
	= a_{21} v_y + 0 + b_2\big(-\tfrac{a_{21}}{b_2} v_y\big) = 0.
	\label{eq:r_branch1_zero_dyn}
\end{align}
Therefore, \(\mathcal{Z}_1\) is an invariant manifold for the closed-loop dynamics: if \(r(0)=0\), then \(r(t)\equiv 0\) for all \(t\), and the zero dynamics reduce to the one-dimensional linear system
\begin{equation}
	\dot{v}_y = a_{11} v_y.
	\label{eq:zd_branch1_zero_dyn}
\end{equation}
For any initial condition \(v_y(0)\neq 0\) with \(r(0)=0\), the solution is
\begin{equation}
	v_y(t) = v_y(0)\, e^{a_{11} t}, \qquad r(t) \equiv 0,
	\label{eq:vy_sol_branch1_zero_dyn}
\end{equation}
and the corresponding input that keeps \(y(t)\equiv 0\) is
\begin{equation}
	u_{z}^{a}(t)=
	= -\frac{a_{21}}{b_2} v_y(0) e^{a_{11} t}
	= \bar{M}_1 e^{a_{11} t},
	\label{eq:M_sol_branch1_zero_dyn}
\end{equation}
where
\[
\bar{M}_1 := -\frac{a_{21}}{b_2} v_y(0).
\]

\textbf{(ii)} On $\mathcal{Z}_2$, we impose $v_y(t) \equiv 0$. Setting $v_y = 0$
in the Lie derivatives yields
\begin{align}
	\dot{y}\big|_{v_y=0} = L_f h(x)\big|_{v_y=0} = -a_{12} r^2,
\end{align}
which is independent of $M_z$ (since $L_g h(x) = -b_2 v_y = 0$). Enforcing
$\dot{y} = 0$ with $a_{12} \neq 0$ requires $r(t) \equiv 0$. With both
$v_y \equiv 0$ and $r \equiv 0$, the state equations reduce to $0 = b_2 M_z$,
which, since $b_2 \neq 0$, forces $M_z(t) \equiv 0$. Hence the only
zero-output trajectory on $\mathcal{Z}_2$ is the equilibrium
$(v_y, r, M_z) = (0, 0, 0)$.

\section{Proof of Proposition 3}
\label{sec:Appendix2}
\textbf{Proof}:
Since $x^a(0) = 0$, the evolution of the attacker's internal model will be $x^a(t) = \int_0^t e^{A(t-\tau)}Bu_c^a(\tau)\,d\tau$. Furthermore, the attacker's goal is to ensure $x(t)=x^n(t)+x^a(t)$ while remaining stealthy. Expanding
$y(t) = -F_1(x^n(t)+x^a(t))F_2(x^n(t)+x^a(t))$ and using
$y^n(t) = -F_1x^n(t)F_2x^n(t)$, we obtain
\begin{align}
	\begin{split}
		y(t) &= y^n(t) - F_1x^a(t)F_2\bigl(x^n(t)+x^a(t)\bigr)
		\\&- F_1\bigl(x^n(t)+x^a(t)\bigr)F_2x^a(t)
	\end{split}
\end{align}
Since $y_c^a(t) = y^n(t) - y(t)$, we can obtain:
\begin{align}
	\begin{split}
		y_c^a(t) &= F_1x^a(t)F_2x^n(t) + F_1x^n(t)F_2x^a(t) \\&+ 2F_1x^a(t)F_2x^a(t)
	\end{split}
\end{align}
Substituting $F_1 = [I,\,0]$ and $F_2 = [0,\,I]$ yields the equation (\ref{eq:CovOutputFinal}). This completes the proof of the theorem. \hfill $\blacksquare$

\bibliographystyle{IEEEtran}
\bibliography{sample}

\begin{IEEEbiographynophoto}{Ali Eslami}
	is a Postdoctoral Researcher at McGill University, Montreal, Canada. He holds a Ph.D. in Electrical Engineering from Concordia University, an M.Sc. from Amirkabir University of Technology, and a B.Sc. from the University of Tehran, both in Iran. His research spans cybersecurity, cyber-physical systems, agentic AI, and resilient control, with a focus on detection and recovery from cyber-attacks in autonomous and connected systems, fault-tolerant control design, and the theoretical foundations of agentic AI for critical infrastructure.
\end{IEEEbiographynophoto} 
\begin{IEEEbiographynophoto}{Jiangbo Yu}
is an Assistant Professor in the Department of Civil Engineering at McGill University, Montreal, Canada, where he leads the Human-Machine Transportation Systems (HMTS) Lab. He holds a Ph.D. in Civil Engineering (Transportation) from the University of California, Irvine, an M.S. from the University of Southern California, and a B.S. from Beijing Institute of Technology. Prior to McGill, he was a Research Associate at MIT and held senior engineering roles at AECOM and Cambridge Systematics. His research focuses on agentic transportation systems, human-AI interaction, automated electric mobility, and infrastructure network resilience. He is a licensed Professional Engineer (P.E.) and a certified Professional Transportation Planner (PTP).
\end{IEEEbiographynophoto}

\begin{IEEEbiographynophoto}{Mohammad Pirani}
	is an assistant professor with the Department of Mechanical Engineering, University of Ottawa, Canada. He was a research assistant professor in the Department of Mechanical and Mechatronics Engineering at the University of Waterloo (2022–2023). He held postdoctoral researcher positions at the University of Toronto (2019–2021) and KTH Royal Institute of Technology, Sweden (2018–2019). He received a MASc degree in electrical and computer engineering and a Ph.D. degree in Mechanical and Mechatronics Engineering, both from the University of Waterloo in 2014 and 2017, respectively. His research interests include resilient and fault-tolerant control, networked control systems, and multi-agent systems.
\end{IEEEbiographynophoto}

\end{document}